\documentclass[%
 twocolumn,amsmath,amssymb,
 aps,
]{revtex4}
\usepackage{graphicx}
\usepackage{dcolumn}
\usepackage{bm}

\usepackage{color}
\newcommand{\rot}[1]{{\color{black} #1}}

\begin{document}

\preprint{APS/number TBD}

\title{Comment on ``Asymptotic Phase for Stochastic Oscillators''}

\author{Peter J.~Thomas}
\affiliation{%
Department of Mathematics, Applied Mathematics, and Statistics.\\
Case Western Reserve University,
Cleveland, Ohio, 44106, USA
}%

\author{Benjamin Lindner}
\affiliation{Bernstein Center for Computational Neuroscience and Department of Physics. \\
Humboldt University, 10115 Berlin, Germany.}

\date{\today}


\begin{abstract}In his Comment [arXiv:1501.02126 (2015)] on our recent paper [Phys.~Rev.~Lett., v.~113, 254101 (2014)], Pikovsky compares two methods for defining the ``phase" of a stochastic oscillator. We reply to his Comment by showing that neither method can unambiguously identify a unique system of isochrons, when multiple oscillations coexist in the same system.
\end{abstract}

\maketitle



In his comment \cite{Pikovsky2015Comment-v1-arxiv} on our paper \cite{LindnerThomas2014PRL}, Pikovsky contrasts two definitions for the phase of a stochastic oscillator by way of an analytically solvable model system. In \cite{SchwabedalPikovsky2013PRL} the phase is defined in terms of a system of isochrons $\Sigma_\theta$, analogous to Poincar\'{e} sections, with the property that for any initial condition on one isochron $\Sigma_{a}$, the mean first passage time (MFPT) to a second isochron $\Sigma_{b}$, $b>a$, will depend only on the phase difference $b-a$. In our approach  \cite{LindnerThomas2014PRL} the phase is defined as the complex argument of the slowest decaying eigenfunction of the backward Kolmogorov operator, provided the first nontrivial eigenvalue is complex and is well separated from the next slowest decaying eigenvalue. Pikovsky argues that our eigenfunction approach does not properly work in all situations and proposes the following example to demonstrate this.  Consider  two independent phase-like variables, each taking values in $[0,2\pi)$, that obey 
\begin{eqnarray}\label{eq:dthetasdt}
\dot{\theta}_1=\omega_1+\sigma_1\xi_1,\hspace{1cm}\dot{\theta}_2=\omega_2+\sigma_2\xi_2,
\end{eqnarray}
where $\langle \xi_i(t)\xi_j(t')\rangle =2\delta_{ij}\delta(t-t')$. 
The eigenvalues of the adjoint Fokker-Planck operator are $\lambda_{n,m}=i(n\omega_1+m\omega_2)-(n\sigma_1)^2-(m\sigma_2)^2$.  The ``slowest decaying mode" corresponds either to $\lambda_{1,0}$  (for $\sigma_1<\sigma_2$) or  to $\lambda_{0,1}$ 
(for $\sigma_1>\sigma_2$).  In the first case, $\theta_1$ could be interpreted as the primary phase variable; in the second case, $\theta_2$ could be.  
\rot{(We adopt the notation $\theta_1,\theta_2$, rather than $\theta,\phi$, so as not to prejudge the roles of the variables.)}
If $\sigma_1\approx\sigma_2$, the system is not ``robustly oscillatory" in the sense of \cite{LindnerThomas2014PRL}, and neither variable is clearly identified as the primary ``phase".

Thus, as  Pikovsky's example illustrates, the spectral method does not unambiguously identify a unique phase variable, when multiple oscillations coexist with similar coherence times. \rot{Pikovsky asserts in his comment that ``the approach of Ref.~\cite{SchwabedalPikovsky2013PRL} yields here the proper phase $\theta=\text{const.}$"}  However, as we demonstrate below, the MFPT method introduced in \cite{SchwabedalPikovsky2013PRL} necessarily exhibits the same ambiguity under the same circumstances, at least for this simple example.  

Let $\psi\in[0,2\pi)$ satisfy $\dot{\psi}=\omega+\sigma\xi$, where $\xi$ is white \rot{Gaussian} noise \rot{and $\omega>0$}.  Let $\tau(a,b)$ be the MFPT of the system starting at $\psi=a$ to arrive at $\psi=b>a$.  This quantity satisfies an equation involving the same backwards operator as that identifying the isochrons in \cite{LindnerThomas2014PRL}, namely 
\begin{equation}
\omega\frac{\partial\tau}{\partial a}+\sigma^2\frac{\partial^2\tau}{\partial a^2}=-1,
\end{equation}
with boundary condition $\tau(a,b)\to 0$ as $a\to b$ \cite{Gardiner2004}.  Clearly the solution is $\tau(a,b)=(b-a)/\omega$.
Therefore the surfaces $\theta_1=\text{const.}$ provide a system of MFPT isochrons for the system \eqref{eq:dthetasdt}, as described in \cite{SchwabedalPikovsky2013PRL}.  However, so do the surfaces $\theta_2=\text{const.}$  Moreover, for any nontrivial pair of integers $(n,m)$, the surfaces $\psi_{n,m}=\text{const.}$ form another system of MFPT isochrons, where we  define
\begin{equation}
\psi_{n,m}=\frac{n\theta_1+m\theta_2}{n+m}.
\end{equation}
\rot{This is easily seen, }
since $\psi_{n,m}$ obeys \rot{a stochastic differential equation (SDE)} of the same form \rot{as $\theta_1$ and $\theta_2$}, 
\begin{equation}
\dot{\psi}_{n,m}=\frac{n\omega_1+m\omega_2}{n+m}+\frac{\sqrt{n^2\sigma_1^2+m^2\sigma_2^2}}{n+m}\xi_3.
\end{equation}
\rot{Therefore,} there is a countably infinite collection of surfaces satisfying the MFPT property for the system he describes.  \rot{It is difficult to see how the MFPT approach would identify a unique system of isochrons, without being supplemented by additional criteria.}

We have omitted the radial variable from our eq.~\eqref{eq:dthetasdt}.  In equations (1-3) of Pikovsky's comment we note that the radial variable is entirely uncoupled from the two phase variables.  The physical motivation for  the example is a noisy limit cycle tracing an orbit in three dimensional space,  rotating simultaneously in both angles describing the points on a torus.   However, neither the SDE nor the Fokker-Planck equation analyzed in the note correspond to this physical system.  What is missing is the interaction of one of the phase variables with the radial variable in the \rot{SDE}.  By decoupling them in the equations given,  Pikovsky has made the system symmetric with respect to exchange of \rot{$\theta_1\leftrightarrow\theta_2$ (equivalently, $\theta\leftrightarrow\phi$)}.  The system lacking this symmetry is more difficult to analyze.  Certainly one could construct  a \rot{3D} system in which the asymptotic phase obtained from the adjoint eigenfunctions appears ambiguous; however in any such system we suspect that the construction based on MFPT isochrons will suffer from the same ambiguity.  

\begin{widetext}
\begin{figure}[t!]
   \centering
   \includegraphics[width=7in]{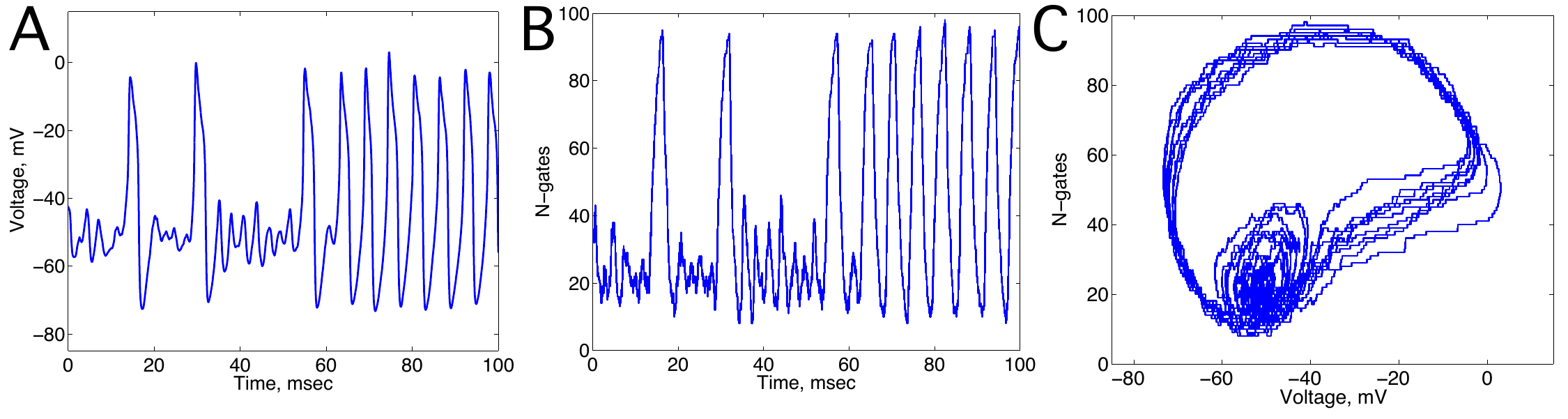}
   \caption{\rot{Multirhythmic behavior in Izhikevich's low-threshold persistent sodium plus potassium model, with parameters giving subcritical Hopf bifurcation near $I\approx 43\mu$A/cm$^2$.  For $N_\text{tot}=100$ discrete, randomly gated potassium channels, and $I=40\mu$A/cm$^2$, trajectories show a mixture of spiking and subthreshold oscillations, at two distinct frequencies. (\textbf{A}) $V$, Voltage (mV) and (\textbf{B}) $N$, number of open potassium channels are plotted against time (msec).  (\textbf{C}) The trajectory in the $(V,N)$ plane shows coexistence of a large and a small oscillation.  For this value of $I$, the eigenvalue spectrum computed in \cite{LindnerThomas2014PRL} shows two slowly decaying complex modes with similar negative real parts. Trajectory generated \textit{via} an exact simulation algorithm \cite{AndersonErmentroutThomas2014JCNS}.}}
   \label{fig:multirhythmic}
\end{figure}
\end{widetext}

Indeed, such ambiguity arises naturally in the case of \rot{multirhythmic} or mixed-mode oscillations.
In \cite{LindnerThomas2014PRL} we analyzed the eigenvalue spectrum of  Izhikevich's ``low-threshold persistent sodium plus potassium" model, with parameters giving a subcritical Andronov-Hopf bifurcation for injected current $I\approx 43\mu$A/cm$^2$ (\cite{Izhikevich2007}, Figs.~6.16 and 4.1b).  Well above the bifurcation point (at $I=60\mu$A/cm$^2$) the eigenvalues follow a nearly parabolic spectrum, as expected for a robustly oscillatory system with weak noise.  \rot{Just below} the bifurcation point \rot{(at $I=40\mu$A/cm$^2$)}, channel noise induces switching between spiking and subthreshold oscillations \rot{(Fig.~\ref{fig:multirhythmic})}.  
The eigenvalue spectrum shows two distinct slowly decaying modes with similar decay rates, reflecting the coexistence of distinct oscillatory processes, each with its own typical frequency.  It is not clear how the MFPT based isochron construction would perform in this setting; in our opinion both approaches merit further development. 

Peter J.~Thomas$^{1,2}$ and Benjamin Lindner$^{2,3}$\\
\small
$^1$Department of Math, Applied Math and Statistics\\
Case Western Reserve University\\
Cleveland, Ohio 44106, USA.\\
$^2$Bernstein Center for Computational Neuroscience\\
10115 Berlin, Germany\\
$^3$Department of Physics,  Humboldt University\\
12489 Berlin, Germany


\onecolumngrid
\twocolumngrid


\begin{thebibliography}{99}
\bibitem{Pikovsky2015Comment-v1-arxiv}
A. Pikovsky. \newblock{arXiv preprint arXiv:1501.02126v2},
2015.
\bibitem{LindnerThomas2014PRL}  P. J. Thomas and B. Lindner 
\newblock{\em Phys. Rev. Lett.}, 113:254101, 
2014.
\bibitem{SchwabedalPikovsky2013PRL} J. Schwabedal and A. Pikovsky  \newblock {\em Phys. Rev. Lett.}, 110:4102, 2013.
\bibitem{Gardiner2004} C. W. Gardiner. {\em Handbook of Stochastic Methods for
Physics, Chemistry, and the Natural Sciences.} Springer
Verlag, 2nd edition, 2004.
\bibitem{Izhikevich2007} E. M. Izhikevich. {\em Dynamical Systems in Neuroscience: The Geometry of Excitability and Bursting.} MIT Press, Cambridge, Massachusetts, 2007.
\bibitem{AndersonErmentroutThomas2014JCNS}
\rot{D. F. Anderson, B. Ermentrout and P. J. Thomas. 
\newblock {\em J. Comput. Neurosci.}, 38(1):67-82, 2014.}
\end{thebibliography}
\end{document}